# Significant enhancement of room temperature ferromagnetism in surfactant coated polycrystalline Mn doped ZnO particles

O. D. Jayakumar, I. K. Gopalakrishnan*, C. Sudakar[#], R. M. Kadam and

S. K. Kulshreshtha

Chemistry Division, Bhabha Atomic Research Centre, Mumbai 400085, India.

[#]Department of Physics and Astronomy, Wayne State University, Detroit, MI 48201, U.S.A.

**Abstract.**

We report a surfactant assisted synthesis of Mn doped ZnO polycrystalline samples showing robust room temperature ferromagnetism as characterized by X-ray diffraction analysis, Transmission Electron Microscopy, Electron Paramagnetic Resonance and *DC* magnetization mearurements. This surfactant assisted synthesis method, developed by us, is found to be highly reproducible. Further, it can also be extended to the synthesis of other transition metal doped ZnO.

Keywords: Mn doped ZnO, dilute magnetic semiconductor oxide materials,

X-ray diffraction, Transmission Electron Microscopy, *DC* magnetization measurements, surfactant assisted synthesis.

*Corresponding author : E-mail: ikgopal@barc.gov.in

# 1. Introduction

There is extensive interest in the growth of wide band gap oxide semiconductors, which are ferromagnetic at room temperature. In particular, Mn doped ZnO has attracted substantial interest [1, 2]. Despite clear theoretical predictions [3, 4], ferromagnetism in Mn doped ZnO remains poorly understood experimentally. Often different laboratories have reported incongruous results for seemingly identical materials. Several authors have observed only paramagnetism at any temperature [5, 6], while others have reported ferromagnetism with $T_C$ below room temperature [7]. Sharma et al. [8] was the first to report room temperature ferromagnetism in thin films by PLD, and bulk powders of Mn doped ZnO, produced by low temperature solid state reaction. They have attributed the observed ferromagnetism to carrier induced, from a theoeretical study. Later, Kundaliya et al. [9] claimed the origin of room temperature ferromagnetism (RTF) to arise from a totally different compound (Zn doped $Mn_2O_3$) and not from Mn substituting for Zn in the ZnO wurtzite structure. However, any doubt about the intrinsic nature of ferromagnetism in Mn doped ZnO was cleared by the more recent work of Kittilstved et al. [10]. They have demonstrated the reversible cycling of paramagnetic (P) to ferromagnetic (RTF) state in Mn doped ZnO nanocrystals by chemical manipulation. However, the situation in the case of bulk materials is still not clear. Hong et al. [11] in a recent paper highlighted the role of defects in tuning ferromagnetism in oxide DMS films. In view of this, we thought it is worth while to explore ways to fine tune synthesis conditions to optimize magnetic characteristics of Mn doped ZnO bulk samples. In this communication, we present our efforts in fine tuning preparation conditions for optimizing ferromagnetic characteristics of Mn doped ZnO bulk samples reliably by surfactant treatment and calcinations, in a reproducible way.

## 2. Experimental

The procedure involved heating a mixture of AOT (sodium bis (2-ethylhexyl) sulfosuccinate) or triblock co-polymer (poly (ethylene glycol)-block-poly (propylene glycol)-block-poly (ethylene glycol)) (Pluronic 123) and the bulk samples of Mn doped ZnO (prepared by the method followed by Sharma et al. [8]) at 400°C for 1 hr, followed by washing and drying of the samples. Dried powders were calcined at 400°C for 4 hours in air. Mn doped ZnO with different Mn concentrations were prepared by standard solid state reaction route [8], using high purity ZnO (99.999%, May & Baker Ltd., Dagenham, UK) and $MnO_2$ (99.9999%, Spex Ind. Inc, N.J., USA) powders. Appropriate amounts of ZnO and $MnO_2$ powders were mixed, calcined at 400°C for 8 hours and sintered in air at 450°C for 12 hours and cooled to room temperature normally to obtain nominal $Zn_{1-x}Mn_xO$ (x = 0.0, 0.01, 0.022, 0.03, 0.04, and 0.05) powders. These samples were then mixed with AOT (sample to AOT weight ratio 1:10), heated at 400°C for 1 hour and subsequently washed using distilled water and absolute ethanol several times and dried at 125°C for 16 hours. This dried powder is annealed at 400°C for 4 hours in air. (The mixing of AOT and Mn doped ZnO sample is accomplished by weighing the required quantity of AOT and making a hollow ball out of it. Then the weighed Mn doped sample is put into the AOT ball. The AOT ball containing the doped sample is then placed in an alumina boat and heated in a tubular furnace at the rate of 10°C/min, at 400°C as mentioned above). Phase purity and the structure of the samples were analyzed using CuKα radiation by employing a Philips

Diffractometer (model PW 1071) fitted with graphite crystal monochromater. The lattice parameters of the compounds were extracted by Rietveld refinement of the XRD data by using the program Fullprof [12] with X-ray intensity collected for the range $10° \leq 2\theta \leq 70°$. *DC* magnetization measurements as a function of field were carried out using an E.G. &G P.A.R vibrating sample magnetometer (model 4500). Electron diffraction and microscopy were carried out with a JEOL, JEM 2010 transmission electron microscope (TEM) for morphological and high-resolution electron microscopy (HREM) studies. Specimens were prepared for TEM by placing a drop of the suspended particles on carbon-coated Cu grids. The EPR experiments were performed using BRUKER ESP 300 spectrometer operated at X-band frequency (9.5 GHz). DPPH was used for calibrating the g values.

## 3. Results and Discussion

Rietveld Profile refinement analysis of XRD data of $Zn_{1-x}Mn_xO$ (x = 0.0, 0.01, 0.022, 0.03, 0.04, 0.05) showed that they are single phase with the wurtzite structure (space group $P6_3mc$). The dependence of cell volume with dopant concentration is presented in Fig.1. The linear increase of cell volume with dopant concentration indicates the incorporation of Mn into ZnO lattice [13, 14]. Since X-ray diffraction analysis being a bulk technique, one cannot detect inclusions of minute secondary phases. In order to make sure the samples are free from secondary phases we have carried out Transmission Electron Microscopic analysis on 2.2 at % Mn doped ZnO treated with AOT.

The bright field images of the particles of sample corresponding to 2.2 at % Mn doped ZnO show narrow distribution in shape and size, with particles exhibiting predominantly rod-like microstructure developing into platelets and sizes in the range of 50 to 200 nm as depicted in Fig.2a. The particles are uniformly coated with the surfactant layer (thickness of coating ~10nm, more clearly seen in Fig.2c). The particles are highly crystalline, homogenous and devoid of any small clusters or secondary phases as has been thoroughly investigated by scanning large portions of the particles on the grid. The selected area electron diffraction (SAED) pattern of one such particle taken with B = [2-1-10] is shown in Fig. 2b. Electron diffraction pattern shows that the particle exhibit monocrystalline wurtzite ZnO structure. The high resolution electron micrograph in Fig. 2c shows the lattice fringe width of ~2.8 Å, corresponding to (01-10) plane of wurtzite ZnO phase.

DC magnetization loops of Mn doped ZnO without AOT is given in figure 3. From this figure it is clear that the increase in Mn concentration decreases the ferromagnetic nature and increases the paramagnetic component of samples in agreement with the results of sharma et al. [8]. Figure 4 shows the *M-H* curves of the same samples after AOT treatment followed by calcination at 400$^o$C. It can be seen that the magnetic moment of all samples, in particular 2.2 at % Mn doped ZnO, increases significantly after AOT treatment followed by 400$^o$C calcinations. For 2.2 at % Mn doped ZnO the saturation magnetization (*Ms*) is 0.05 emu/g with a coercieve field (*Hc*) of 125 Oe (inset Fig.5) and remanence (*Mr* ) of 0.008 emu/g. This value of *Ms* is about 7 times the value (0.007 emu/g) obtained by sharma et al. [8] for 2.2 at % Mn doped ZnO, and about 5 times the value (0.01 emu/g) of our 2.2 % Mn doped as synthesized sample. The decrease of ferromagnetic nature of samples with increase in Mn concentration beyond 2.2 at % Mn may be due to the clustering as a consequence of decreased Mn-Mn distance leading to antiferromagnetic nearest neighbour exchanges [8]. Sharma et al. [8] observed maximum ferromagnetic moment for their 2.2 at % Mn doped ZnO and found that the ferromagnetic nature decreases with increase in Mn concentration. In our case also, we have observed maximum ferromagnetic moment for 2.2 at % Mn doped sample and so we have given more emphasize to this particular sample. To see the effect of different surfactants on the magnetic properties, we treated the sample 2.2 at % Mn doped ZnO using tri block co-polymer (Pluronic 123) instead of AOT. The results are presented in Fig.5. It can be seen that the AOT coated sample shows almost 100% enhancement in saturation magnetization value as compared to co-polymer treated sample. These experiments, for all Mn compositions, were repeated several times using both AOT and co-polymer and have been found to be highly reproducible.

Fig.6 shows the EPR spectra recorded at room temperature on 2.2 at % Mn doped ZnO with and without AOT coating. EPR spectrum of 2.2 at % Mn doped samples clearly showed two signals one at low field corresponding to the FMR signal [8,15,17] which arises from transition within the ground state of ferromagnetic domain, and the other due to the well known EPR spectrum (sextet hyperfine structure) [16] of $Mn^{2+}$ ions which may not be participating in the ferromagnetic ordering. It can be seen from the EPR spectrum (Fig.6), that for sample which is not treated with AOT, the signal corresponding to $Mn^{2+}$ ions is more intense than that treated with AOT. Further, for this sample, the FMR signal intensity is much subdued as compared to AOT treated sample. This suggests that a relatively large concentration of manganese in uncoated sample is not participating in the long range ferromagnetic ordering. This may be the reason for the observed low value of saturation magnetization M$s$ (0.01 emu/g) observed for the uncoated sample than the AOT coated sample ($Ms = 0.05$ emu/g). Sharma et al. [8] observed a saturation moment of 0.007 emu $g^{-1}$ for bulk Mn doped ZnO prepared by solid state reaction between ZnO and $MnO_2$ where they observed less than 4 % manganese was responsible for ferromagnetism at room temperature. In the present study we observed saturation magnetic moments of $Ms = 0.05$ emu $g^{-1}$ for AOT coated 2.2 at % Mn doped polycrystalline samples at RT, which is significantly greater than the bulk Mn doped ZnO [8].

It is well documented that the presence of surfactant reduces interfacial tension between the host and the dopant [19-21]. The presence of surfactant may inhibit the clustering of the dopant ions (Mn) and make the Mn doped ZnO more homogeneous. This facilitates the suppression of antiferromagnetic exchange interaction between the nearest neighbour dopant ions and the enhancement of ferromagnetic exchange interaction among the widely separated dopant ions mediated through defect/carriers generated by the surfactant treatment. Thus the observed enhancement of ferromagnetism upon treatment of surfactant can be attributed to the homogenous distribution of this dopant ions and the fusion of defects at interfaces between the crystallites. Both HREM images and SAED pattern of AOT coated sample presented in Fig.2 also support this conjecture. Similar results was observed by Gamelin`s group [10, 22, 23]. Their work unambiguously demonstrated the role of interfacial or grain-boundary defects in activating room temperature ferromagnetism in a broad spectrum of transition metal doped oxides including cobalt-doped ZnO. Ramchandran et al. [24] studying microstructure of Mn doped ZnO films have shown that a close correlation exists between microstructural features and the appearance of room temperature ferromagnetism. Our results also can explain the preparation dependent ferromagnetism observed by different groups in Zn-Mn-O system. We feel this surfactant coated enhancement of magnetization developed by us will help to resolve many ambiguities surrounding the origin of ferromagnetism in the Zn-Mn-O system. This method of enhancement of magnetic moment can also be extended to other transition metal doped ZnO. Further studies are in progress to find out the exact cause for the enhancement of magnetic moment and also to further improve the magnetic characteristics.

**4. Conclusions**

Mn doped ZnO polycrystalline samples showing robust room temperature ferromagnetism, have been synthezised by a surfactant assisted method. X-ray diffraction analysis showed an impurity free single phase compound formation. HRTEM and SAED studies of Mn doped ZnO revealed that the particles are highly monocrystalline, homogenous and devoid of any small clusters or secondary phases with wurtzite type ZnO structure. FMR signal observed at RT further confirmed the ferromagnetic nature of AOT treated Mn 2.2 % : ZnO. This synthetic procedure allows the relatively rapid preparation of gram quantities of Mn doped ZnO in open containers without any need for special precautions to avoid atmospheric contamination, or temperature stabilization. This method can also be extended to other transition metal doped ZnO.

## Acknowledgments

Authors are grateful to Prof K.V. Rao of Royal Institute of Technology, Stockholm, Sweden for many helpful discussions during the period of this work.

# References

[1] W. Prelleier, A. Fouchet and B. Mercey, J. Phys.:Condens. Matter. 15 (2003) 1583, and references there in.

[2] T. Fukumura, H. Toyosaki and Y. Yamada, Semicond.Sci.Technology 20 (2005) 103, and references there in.

[3] T. Dietl, H. Ohno, F. Matsukura, J. Cibert, and D. Ferrand, Science 287 (2000) 1019.

[4] K. Sato and H. Katayama-Yoshida Jpn.J.Appl.Phys. 39 (2000) L555.

[5] S. S. Kim, J. H. Moon , B.T. Lee, O. S. Song and J. H. Je , J.Appl. Phys. 95 (2004) 454.

[6] A. Tiwari, C. Jin, A. Kvit, D. Kumar, J. F. Muth and J. Narayan Solid State Commun. 121 (2002) 371.

[7] S. W. Jung, S. J. An, G. C. Yi, C. U. Jung, S. I. Lee and S. Cho Appl. Phys. Lett. 80 (2002) 4561.

[8] P. Sharma, A. Gupta, K. V. Rao, F. J. Owens, R. Sharma, R. Ahuja, J. M. O. Gillen, B. Johasson and G. A. Gehring Nature Materials 2 (2003) 673.

[9] D. C. Kundaliya, S. B. Ogale, S. E. Lofland, S. Dhar, C. J. Metting, S. R. Shinde, Z. Ma, B. Varughese, K. V. Ramanujachary, R. L. Salamanca, and T Venkatesan, Nature Materials 3 (2004) 709.

[10] K. R. Kittilstved, N. S. Norberg and D. R. Gamelin, Phys. Rev. Lett. 94 (2005) 147209.

[11] N. H. Hong, J. Sakai, N. T. Huong, N. Poirot, and A. Ruyter Phys. Rev. B, 72 (2005) 045336.

[12] J. Rodriguez-Carvajal, Fullprof: a program for Rietveld Refinement and Profile Matching analysis of Complex Powder Diffraction Patterns ILL.

[13] C. Y. Qin, L. X. Hui, X. U. X. Yu, LI. Lin,. C. J. Ping, W. R. Ming, and D. Peng, Chin. Phys. Lett. 20 (2003) 2058.

[14] S. Kolesnik and B. Dabrowski J. Apply. Phys. 96 (2004) 5379.

[15] N. S. Norberg, K. R. Kittilstved, J. E. Amonette, R. K. Kukkadapu, D. A. Schwartz and D. R. Gamelin, J. Am. Chem. Soc., 126 (2004) 9387.

[16] H. Zhou, D. M. Hofmann, A. Hofstaetter and B. K. Meyer, Journal of Apply. Phys., 94 (2003) 1965.

[17] O. D. Jayakumar, H. G. Salunke, R. M. Kadam, M. Mohapatra, G. Yaswant and S. K. Kulshreshtha, Nanotechnology, 17 (2006) 1278.

[18] J.A. Weil, J.R. Bolton, J E Wertz, Electron Paramagnetic Resonance Elementary Theory and Practical Applications (Wiley, New York 1994) p 498.

[19] M. Copel., M. C. Reuter, E. Kaxiras and R. M. Tomp, Phys.Rev.Lett. 63 (1989) 632.

[20] W. Zhu, H. H. Weitering, E. G. Wang, E. Kaxiras and Z. Zhang, Phys.Rev.Lett. 93 (2004) 126102.

[21] S. C. Erwin, L. Zu, M. I. Haftel, A. L. Efros, T. A. Kennedy and D. J. Norris, Nature 436 (2005) 91.

[22] D. A Schwartz, N. S. Norberg, Q. P Nguyen, J. M Parker and D. R. Gamelin,. J. Am. Chem. Soc., 125 (2003) 13205.

[23] D. A Schwartz and D. R. Gamelin,. Adv. Mater. 16 (2004) 2115.

[24] S. Ramachandran, J. Narayan and J. T. Prater, Appl.Phys. Lett. 88 (2006) 242503.

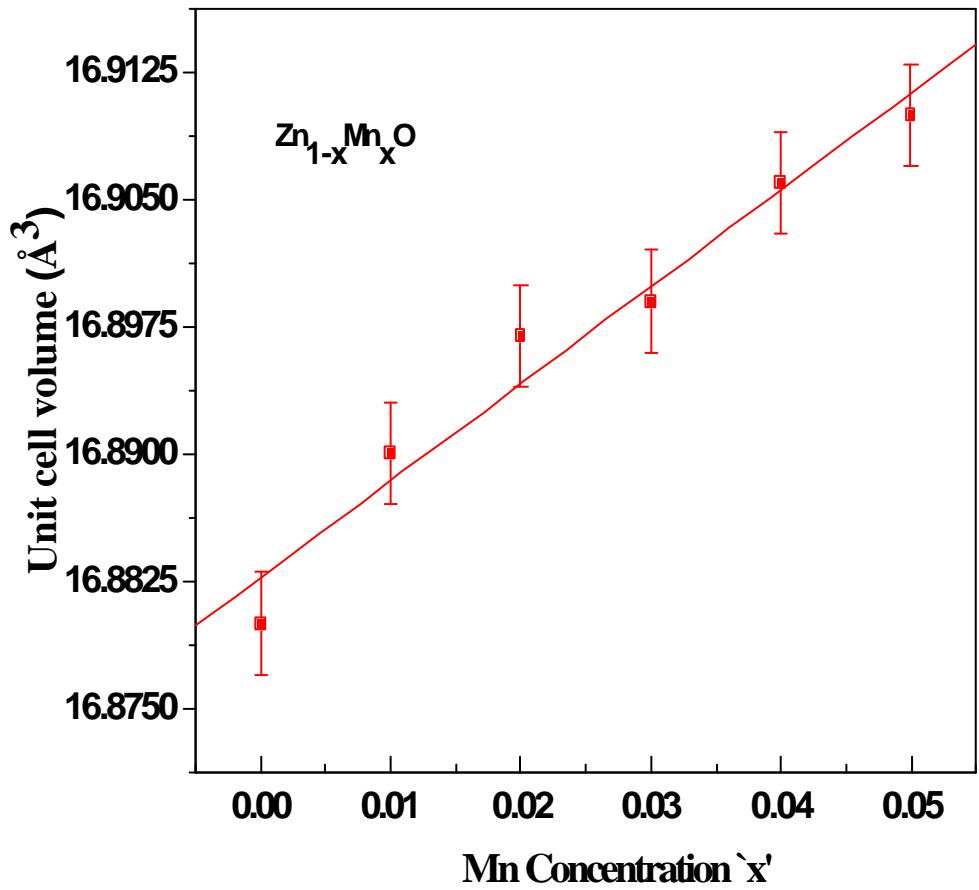

Figure 1. Cell volume dependence on dopant concentration *'x'* in $Zn_{1-x}Mn_xO$.

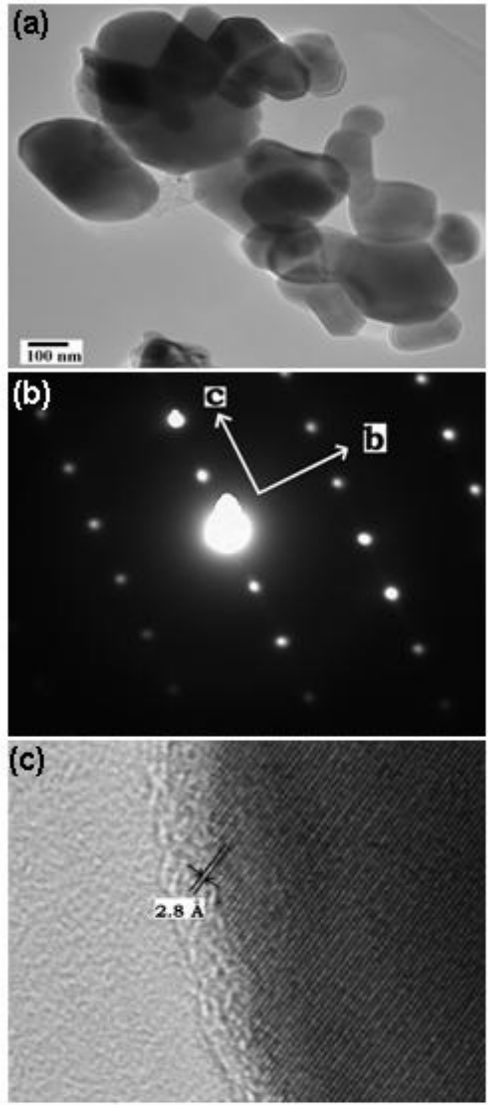

Figure 2 a) TEM of 2.2 at % Mn doped ZnO treated with AOT b) Select area electron diffraction pattern of sample treated with AOT c) HRTEM of 2.2 at % Mn doped ZnO treated with AOT.

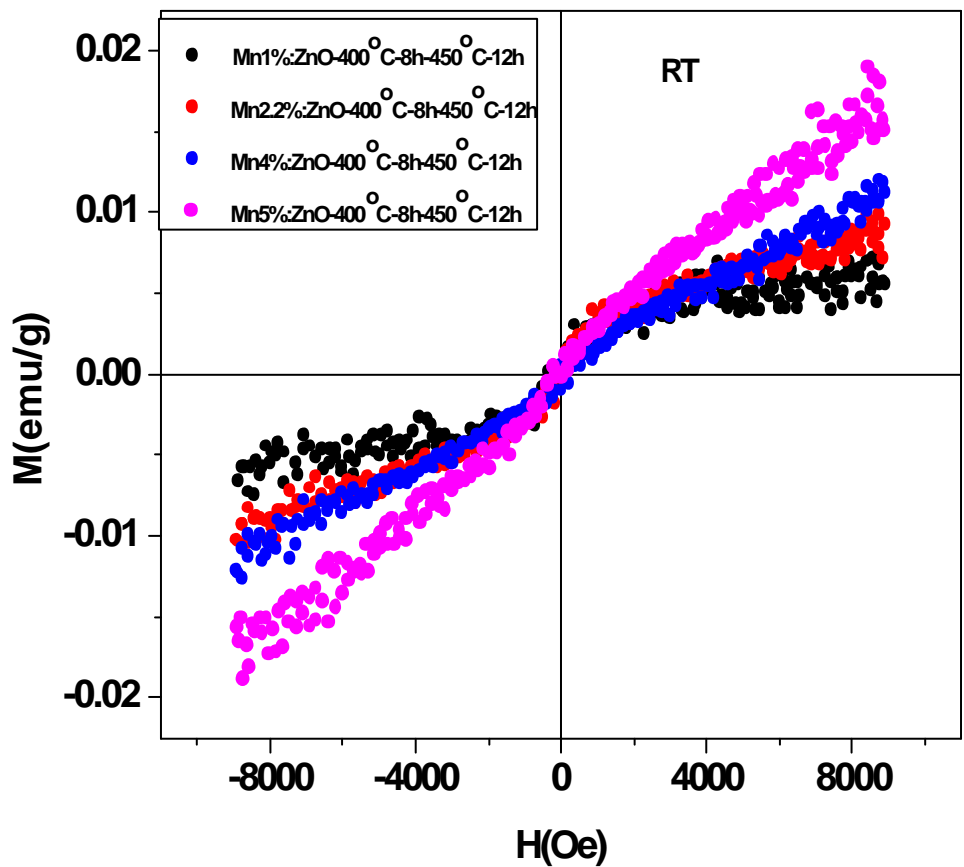

Figure 3. M vs. H curves measured at RT for as prepared samples of $Zn_{1-x}Mn_xO$ (x= 0.01, 0.022, 0.04 and 0.05).

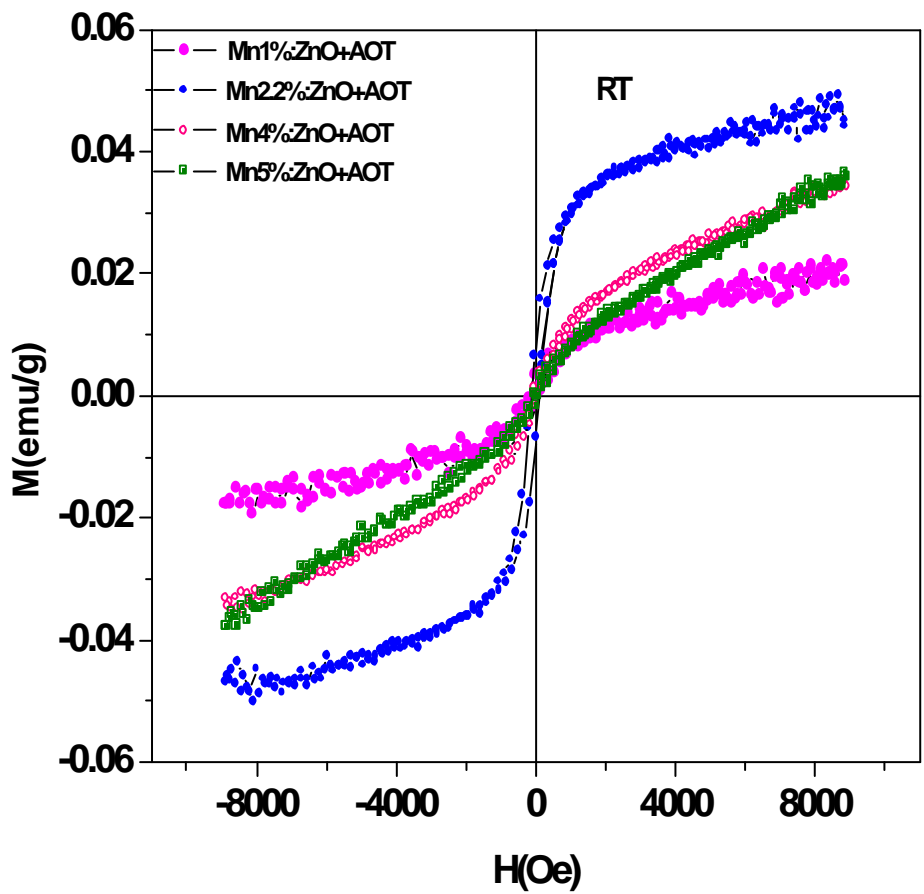

Figure 4. M vs. H curves measured at RT for AOT treated samples of $Zn_{1-x}Mn_xO$ ( x= 0.01, 0.022, 0.04 and 0.05).

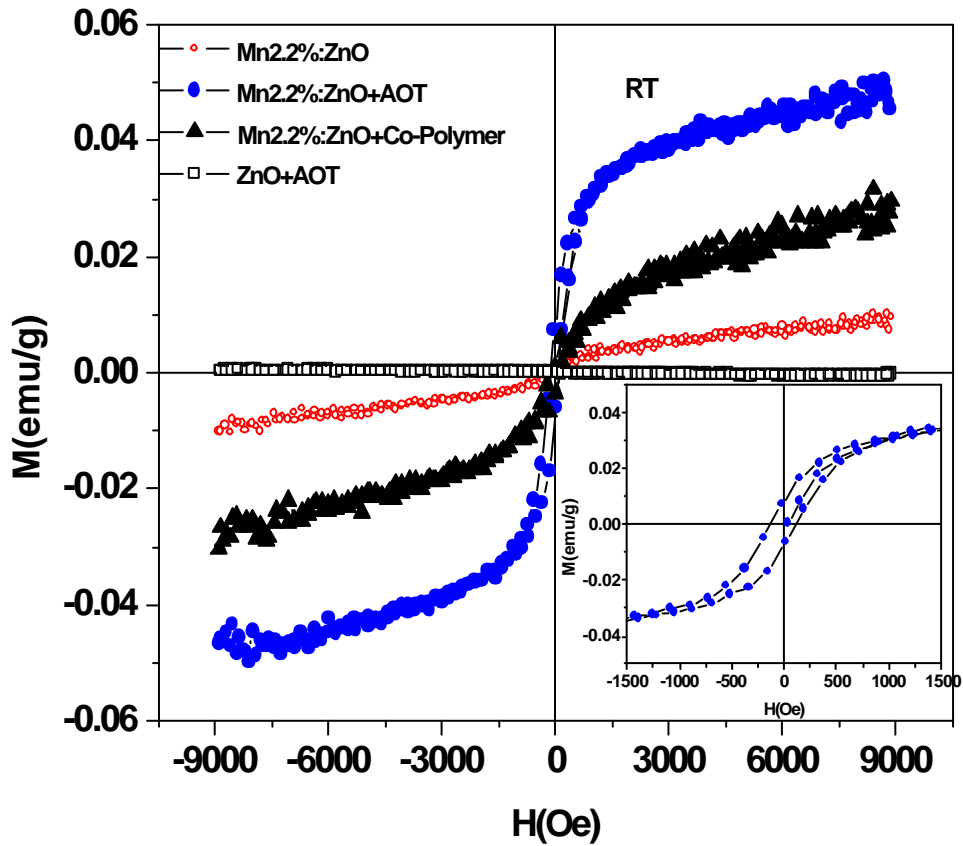

Figure 5. M vs. H curves measured at RT for 2.2 at % Mn doped ZnO samples with and without surfactant treatments. Inset shows data on AOT treated sample in an enlarged scale around origin.

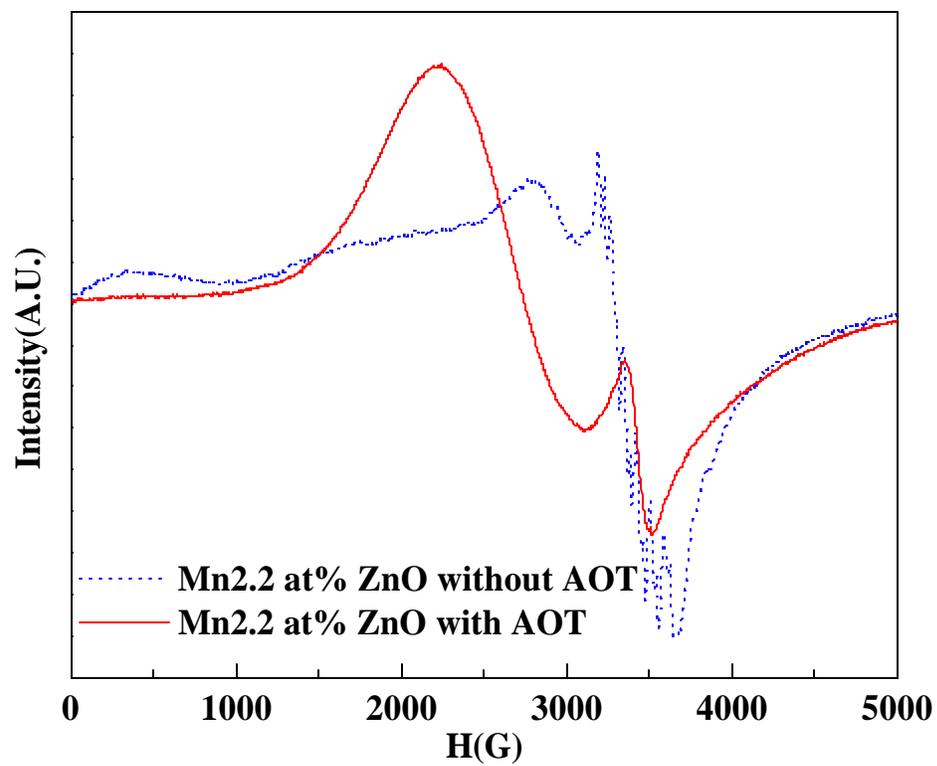

Figure 6. EPR spectra recorded at RT for 2.2 at % Mn doped ZnO with and without AOT coating.